\documentclass[%
 reprint,
 amsmath,amssymb,
 aps,
 a4paper,
 dvipdfmx
]{revtex4-1}

\usepackage[dvipdfmx]{graphicx, color}
\usepackage{dcolumn}
\usepackage{bm}
\usepackage{xcolor}
\usepackage{ulem}
\usepackage{comment}

\newcommand{\bu}{\boldsymbol{u}}
\newcommand{\dfracp}[2]{\dfrac{\partial #1}{\partial #2}}
\newcommand{\dfracd}[2]{\dfrac{{\rm d} #1}{{\rm d} #2}}
\newcommand{\wt}{\widetilde}

\newcommand{\ave}[1]{\left\langle #1 \right\rangle}

\begin{document}

\title{Universality of discontinuous bifurcations in collisionless dynamics}
   \author{Yoshiyuki Y. Yamaguchi$^{1}$}
 \email{yyama@amp.i.kyoto-u.ac.jp}
 \author{Julien Barr{\'e}$^{2}$}
 \email{julien.barre@univ-orleans.fr}
 \affiliation{
   $^{1}$Graduate School of Informatics, Kyoto University, Kyoto 606-8501, Japan\\
   $^{2}$Institut Denis Poisson, Universit{\'e} d'Orl{\'e}ans, Universit{\'e} de Tours and CNRS, 45067 Orl{\'e}ans, France}

\begin{abstract}

  We investigate the universality in collisionless nonlinear dynamics of a codimension-two bifurcation where two eigenvalues collide at the origin, and two lines of continuous bifurcation and discontinuous jump meet. Through linear analysis and direct numerical simulations, we show that this bifurcation does occur, both for two-dimensional shear flows and for repulsive systems mimicking plasmas.

\end{abstract}
\maketitle

\section{Introduction}

Collisionless kinetic equations of Vlasov type typically describe the continuous limit of Hamiltonian systems of particles interacting at long range. As such, they appear in many
different fields of physics; plasma physics and astrophysics are two obvious examples, but there are actually many others: two-dimensional ideal fluids (Euler equation is formally very similar to Vlasov equation \cite{Chavanis96}), non linear optics (see for instance the review \cite{Picozzi14}, section 2), bubbly fluids \cite{Smereka02}, and Free Electron Lasers physics 
(the particles-wave interaction model developed in \cite{Bonifacio84} would be described at the continuous level by an appropriate Vlasov equation).

\begin{figure}
    \includegraphics[width=8cm]{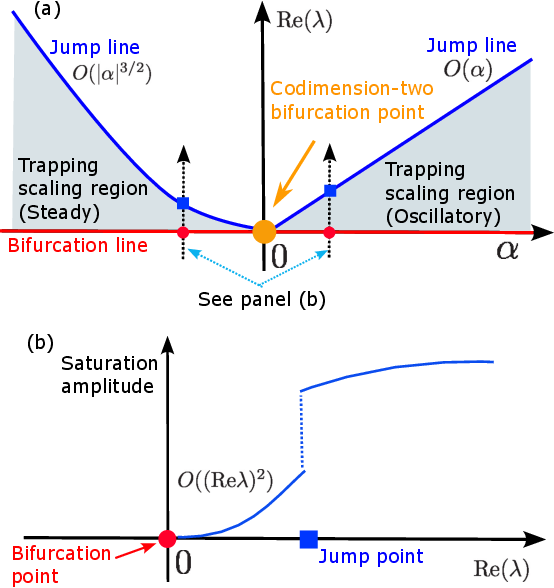}
  \caption{
    (a) Sketch of the two-dimensional parameter space as found in \cite{Yamaguchi-Barre-23}
    around the codimension-two bifurcation point.
    The parameters are ${\rm Re}(\lambda)$, where $\lambda$ is the most unstable eigenvalue,
    and the flatness $\alpha$ of the reference state
    (see the text for the definition).
    The negative side $\alpha<0$ induces a steady bifurcation
    (the trapped state is steady),
    and the positive side $\alpha>0$ an oscillatory bifurcation
    (the trapped state is oscillatory).
    The bifurcation line ${\rm Re}(\lambda)=0$ meets the jump line
    at the codimension-two point only.
    The jump line is non analytic at the codimension-two point, with an exponent $1$ on the right,
    and
    $3/2$ on the left \cite{Yamaguchi-Barre-23}.
    (b) Sketch of a curve representing 
    the saturation amplitude of the instability as a function of ${\rm Re}(\lambda)$,
    along the dashed vertical lines on panel (a).
    The bifurcation is continuous with the trapping scaling $O(({\rm Re}(\lambda))^{2})$,
      except for the codimension-two bifurcation point,
    and the asymptotic amplitude then shows a jump for larger ${\rm Re}(\lambda)$.
  }
  \label{fig:intro} 
\end{figure}

These equations typically have many stationary states. The stability of these states, as well as the non linear behavior of instabilities, are then natural questions of interest. These are bifurcation problems, and, given the common mathematical structure of the underlying equations, one may hope to reach generic results, valid uniformly across the different physical fields considered.  
Investigating bifurcations in Vlasov equation indeed has a long and rich history in plasma physics, and is still an active subject, see for instance  
\cite{ONeil71,Dewar73,Crawford94,Crawford95,Lancellotti03,Tacu22}. It is also an old and active topic in the astrophysical literature (see for instance \cite{Morozov80,Palmer90,Kaur18,Sellwood22,Hamilton24}) and for two-dimensional fluids, where it is closely linked with "critical layer theory" and often regularized by a small viscosity, see for instance \cite{Huerre87,Churilov87,Balmforth97}.
Among the conclusions of these studies, let us stress an important point: resonances between the growing unstable mode and some particles, which are mathematically described by a continuous spectrum, play an important role, and typically make these bifurcations very different from the standard bifurcations of dissipative systems. 
As a major result, these efforts eventually led to the unified understanding of a generic continuous bifurcation for Vlasov-like systems \cite{delCastilloNegrete98,delCastilloNegrete98b,Balmforth13}
through the "Single Wave Model" \cite{Tennyson94,ElskensBook}.
It is characterized by i) "trapping scaling": the saturation of the unstable mode amplitude scales as the square of the instability rate (as opposed to its square root for a standard pitchfork or Hopf bifurcation for instance), and ii) a reduced dynamics described by the coupling of the unstable mode with the resonant particles close to the instability threshold.
This bifurcation is generic, but there exist other types of continuous bifurcations,
when resonance is strong \cite{Crawford96,Balmforth02}, or when it is weak or absent \cite{Barre-Metivier-Yamaguchi-16,Barre-Metivier-Yamaguchi-20}.
It is also known that bifurcations can be discontinuous \cite{Antoniazzi07}, i.e. the saturated amplitude of the unstable mode shows a jump
at the instability threshold; this behavior is sometimes termed "subcritical". In particular, discontinuous bifurcations have been found when the stationary state has a flat top shape (Sec.~\ref{sec:repulsive-case} gives a precise definition), see \cite{Balmforth12} and \cite{Balmforth13} section 8. In \cite{Yamaguchi-Barre-23}, it has been shown that the appearance of a discontinuous bifurcation is associated to the presence of a codimension-two bifurcation, reached for a flat top distribution. This codimension-two point is characterized at the linear level by the collision of two eigenvalues precisely at the origin. At the non linear level, the dynamics in the neighborhood of this point is quite peculiar: in the two-dimensional space of parameters, there is a curve of continuous bifurcations which meets a curve of jump only at the codimension-two point; the jump curve is singular at the codimension-two point. These findings are summarized on Fig.\ref{fig:intro}.    

As made clear from Fig.\ref{fig:intro}, the codimension-two point may shape the parameter space well beyond its immediate neighborhood, making it an important feature for the qualitative description of Vlasov bifurcations. However,
the study in \cite{Yamaguchi-Barre-23} is limited to Vlasov equations describing simple one-dimensional attractive models. In the present work, we thus address the natural questions: 
(i) How general is this codimension-two bifurcation? 
(ii) Can the phenomenology highlighted in \cite{Yamaguchi-Barre-23}, with a collision of eigenvalues at the origin and the non linear behavior described by Fig. \ref{fig:intro}, be found in other models?

We answer these questions by computing eigenvalue diagrams and performing precise numerical simulations for a shear flow in a two dimensional Euler fluid, and for simple repulsive models (thus mimicking a plasma rather than a self-gravitating system). For the two dimensional Euler equation, we conclude that the scenario of \cite{Yamaguchi-Barre-23} summarized in Fig.\ref{fig:intro} is fully valid; we are even able to compute the exponents $1$ and 
$3/2$ characterizing the singularity of the jump curve around the codimension-two point.
For the repulsive model, we do confirm the eigenvalue diagram,
and the existence of a jump at the codimension-two point. The jump is very small however, making further numerical analysis very demanding.
An explanation for the smallness of the jump is also discussed.

This paper is organized as follows.
The Euler fluid is studied in Sec.~\ref{sec:EulerFluid},
and the repulsive case in Sec.~\ref{sec:repulsive-case}.
In each of the two sections, we perform the linear analysis to obtain eigenvalues
and numerical tests to produce bifurcation diagrams.
The last section \ref{sec:summary} is devoted to a summary.

\section{Euler fluid}
\label{sec:EulerFluid}

\subsection{Model}
\label{sec:Model}

We consider the two-dimensional Euler equation on a two-dimensional torus $\mathbb{T}^{2}$,
\begin{equation}
  \label{eq:Euler}
  \dfracp{\omega}{t} + \bu\cdot\nabla\omega = 0,
\end{equation}
where the vorticity field $\omega$ and the velocity field $\bu=(u,v)$
are related to a stream function $\psi$ through
\begin{equation}
  \omega = - \nabla^{2}\psi,
  \quad
  u = \dfracp{\psi}{y},
  \quad
  v = - \dfracp{\psi}{x}.
\end{equation}
The periods for the $x$ and $y$ directions are set as $2\pi L_{x}$ and $2\pi L_{y}$ respectively,
and from now on we fix $L_{x}=1$.
The initial base stream function, representing a shear flow, is set as
\begin{equation}
  \psi_{0}(y) = \cos\dfrac{y}{L_{y}} - a_{2} \cos\dfrac{2y}{L_{y}} - a_{3} \cos\dfrac{3y}{L_{y}},
\end{equation}
and the corresponding vorticity is denoted by
\begin{equation}
  \omega_{0}(y) = \dfrac{1}{L_{y}^{2}} \left(
    \cos\dfrac{y}{L_{y}} - 4 a_{2} \cos\dfrac{2y}{L_{y}} - 9 a_{3} \cos\dfrac{3y}{L_{y}}
  \right),
\end{equation}
where $a_{2},a_{3}\geq 0$. We assume that 
$a_{2}$ and $a_{3}$ are sufficiently small
so that the velocity field $u_{0}(y)=\psi_{0}'(y)$ vanishes
only at $y=0$ and $y=\pi L_{y}$. The explicit condition is
\begin{equation}
  4a_{2} + 9a_{3} < 1.
\end{equation}
The vorticity $\omega$ is analogous to the one-body distribution function $F$ of the Vlasov equation,
and the shear profile $\omega_{0}(y)$ is stationary, as is a spatially homogeneous distribution $F_{0}(p)$ in the Vlasov equation.

\subsection{Linear Analysis}
\label{sec:Linear-Analysis}

\begin{figure}[hbp]
  \centering
    \includegraphics[width=8cm]{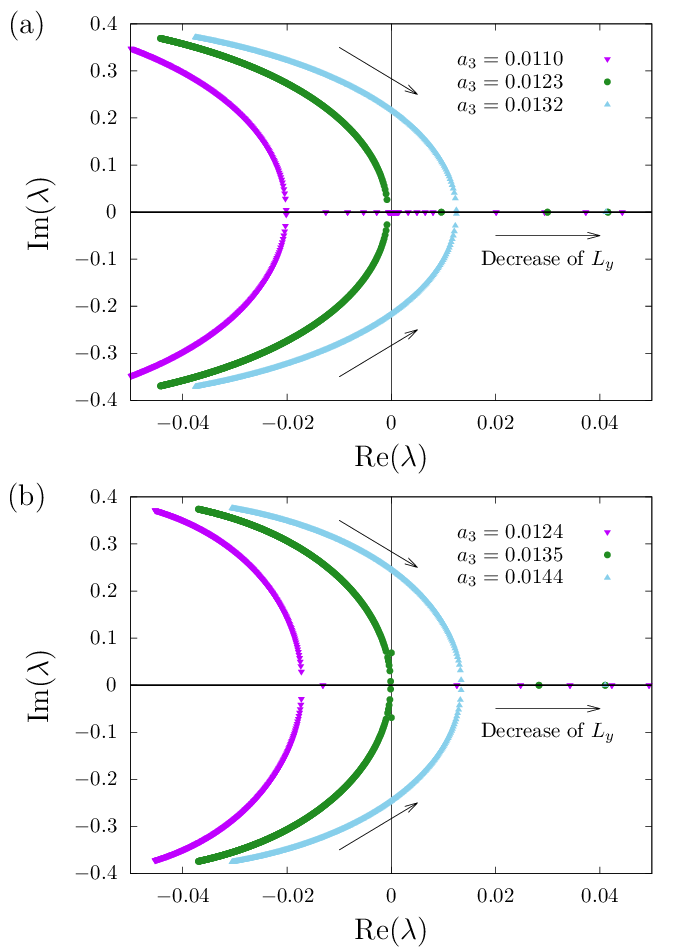}
  \caption{
    Movement of eigenvalues on the complex $\lambda$ plane varying $L_{y}$.
    (a) $a_{2}=0.00$. (b) $a_{2}=0.02$.
    $L_{y}$ decreases from left to right.
  }
  \label{fig:FluidEigenvalue}
\end{figure}

\begin{figure}[hb]
  \centering
    \includegraphics[width=8cm]{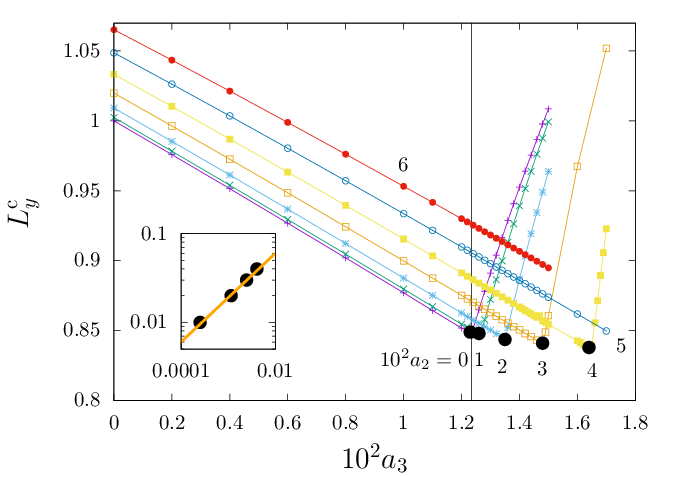}
  \caption{Critical point $L_{y}^{\rm c}$ as a function of $a_{3}$.
    The numbers in the panel represent $10^{2}a_{2}$.
    The minimum (nonsmooth) points (black circles) corresponds
    to the eigenvalue collision at the origin.
    The vertical black line marks $a_{3}=1/81$.
    Inset: Points of collision at the origin on $(a_{3}-1/81,a_{2})$ plane in logarithmic scale.
    The orange line is a guide for $a_{2}=O(\sqrt{a_{3}-1/81})$.
  }
  \label{fig:FluidCriticalPoint}
\end{figure}

The linear stability of $\omega_{0}(y)$ is a classical but delicate problem, as it involves analytical continuations in order to compute the "quasimodes", analogues of Landau poles \cite{Briggs70,Spencer97}.
In the following, for simplicity we will use the term ``eigenvalue''
to denote both true eigenvalues and Landau poles.
We first briefly sketch how to obtain the continued spectrum function,
whose roots are eigenvalues or quasimodes (Landau poles).

The expansion $\psi(x,y,t)=\psi_{0}(y)+\psi_{1}(x,y,t)$ provides
the linearized Euler equation
\begin{equation}
  \dfracp{\omega_{1}}{t} + u_{0}(y) \dfracp{\omega_{1}}{x} + u_{0}''(y) \dfracp{\psi_{1}}{x} = 0.
\end{equation}
We expand $\psi_{1}(x,y,t)$ into a Fourier series with respect to $x$.
Eigenvalues or Landau poles denoted by $\lambda$ are obtained for each Fourier mode $k$
as roots of the continued spectrum function $D_{k}(\lambda)$,
which is computed from solutions of the homogeneous Rayleigh equation
\begin{equation}
  \dfracd{{}^{2}\phi}{y^{2}} - q(y) \phi = 0,
  \quad
  q(y) = k^{2} + \dfrac{u''(y)}{u(y)-i\lambda/k}.
\end{equation}
Let $\phi_{1}$ and $\phi_{2}$ be solutions with boundary conditions
\begin{equation}
  \phi_{1}(0) = 1, \quad \phi_{1}'(0) = 0,
  \quad
  \phi_{2}(0) = 0, \quad \phi_{2}'(0) = 1.
\end{equation}
The spectrum function $D_{k}(\lambda)$ is defined by
\begin{equation}
  D_{k}(\lambda) = \det
  \begin{pmatrix}
    \phi_{1}(2\pi L_{y}) - 1 & \phi_{2}(2\pi L_{y}) \\
    \phi_{1}'(2\pi L_{y}) & \phi_{2}'(2\pi L_{y}) - 1 \\
  \end{pmatrix}.
\end{equation}
The integral contour to solve the Rayleigh equation is on the real axis of $y$
for ${\rm Re}(\lambda)>0$,
but it is continuously modified in the complex $y$ plane
to avoid the singularities in $q(y)$ for ${\rm Re}(\lambda)\leq 0$.
This modification corresponds to the analytical continuation,
and we call $D_{k}(\lambda)$ the continued spectrum function.
See \cite{OBMY-14} for details on the analytical and numerical procedures.
  
In the Vlasov case \cite{Yamaguchi-Barre-23},
assuming that a homogeneous momentum distribution $F_{0}(p)$ is even,
the  codimension-two bifurcation corresponds to a collision of two eigenvalues at the origin.
This collision occurs if and only if
the maximum of $F_{0}(p)$ is flat: $F_{0}''(0)=0$
(see Appendix \ref{sec:collision-flatness} for a proof).
In the 2D Euler case
the analog of the momentum distribution $F_{0}(p)$ is the vorticity profile $\omega_{0}(y)$,
and it has two extrema at $y=0$ and $y=\pi L_{y}$.
The second derivative
\begin{equation}
  \omega_{0}''(y) = - \dfrac{1}{L_{y}^{4}} \left(
    \cos\dfrac{y}{L_{y}} - 16 a_{2} \cos\dfrac{2y}{L_{y}} - 81 a_{3} \cos\dfrac{3y}{L_{y}}
  \right)
\end{equation}
vanishes simultaneously at the two extrema if and only if $(a_{2},a_{3})=(0,1/81)$.
And indeed, as expected, for these parameters an eigenvalue collision at the origin occurs
[see the green points in Fig.~\ref{fig:FluidEigenvalue}(a)].
However, as reported in Fig.~\ref{fig:FluidEigenvalue}(b) and Fig.~\ref{fig:FluidCriticalPoint},
an eigenvalue collision at the origin actually occurs
for a one-dimensional set of $(a_2,a_3)$ parameters:
this is in line with the fact that a codimension-two bifurcation can be found by tuning just two parameters, here $L_y$ and, say, $a_3$; it also shows that the eigenvalue collision at the origin is not strictly related to the flatness of the vorticity profile, which is a remarkable difference with the previous Vlasov case.   
Clearly, the collision at the origin occurs at a critical point, where the real part of the eigenvalue is zero.
Figure \ref{fig:FluidCriticalPoint} shows the critical length $L_{y}^{\rm c}$ as a function of $a_3$, at fixed $a_2$.
Each of the curves has a nonsmooth point, which corresponds to the eigenvalue collision at the origin, illustrating that there is a 
whole family of codimension-two bifurcations.
We denote $a_{3}$ at the nonsmooth point as $a_{3}^{\rm col}$ for  fixed $a_{2}$; $\Delta a_{3}=a_3-a_{3}^{\rm col}$ thus plays the role of the $\alpha$ parameter of Fig.~\ref{fig:intro}. This nonsmooth point at the codimension-two bifurcation point was also found in \cite{Yamaguchi-Barre-23}, suggesting it is a general feature.

\subsection{Numerical tests}

We use a semi-Lagrangian algorithm to simulate the 2D Euler system.
The phase space $(x,y)\in [0,2\pi L_{x})\times [0,2\pi L_{y})$ is divided into an $M\times M$ mesh,
and the temporal backward evolution of a mesh point is computed using
a second-order Runge-Kutta method with time step $\Delta t=0.01$.
A bicubic interpolation is used to estimate the value of $\omega(x,y)$ at a backwards evolved point.
Details on the procedure are given in the Appendix \ref{sec:semi-Lagrangian}.

\begin{figure}[h]
  \centering
      \includegraphics[width=8cm]{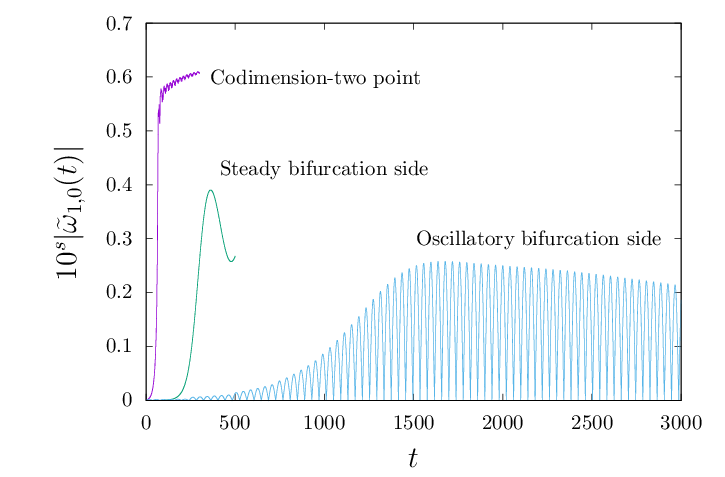}
    \caption{
      Examples of temporal evolution of $|\wt{\omega}_{1,0}(t)|$ in three regions. $a_{2}=0$.
      (Codimension-two point) $a_{3}=0.0123$ and $L_{y}=0.840$. $\epsilon=10^{-3}$. $s=0$.
      (Steady bifurcation side) $a_{3}=0.011$ and $L_{y}=0.8612$. $\epsilon=10^{-6}$. $s=1$.
      (Oscillatory bifurcation side) $a_{3}=0.0126$ and $L_{y}=0.840$. $\epsilon=10^{-6}$. $s=3$.
      The mesh size is $M=256$.
  }
  \label{fig:FluidExample}
\end{figure}

The base stationary flow $\psi_{0}(y)$ is perturbed as
\begin{equation}
  \psi_{0}^{\epsilon}(x,y) = \psi_{0}(y) + \epsilon \cos\dfrac{x}{L_{x}},
\end{equation}
where the initial perturbation amplitude $\epsilon$ ranges from $10^{-3}$ to $10^{-6}$.
The emergence of a spatial pattern in the $x$ direction is observed through the Fourier coefficient
\begin{equation}
  \wt{\omega}_{1,0}(t)
  = \dfrac{1}{2\pi L_{x}} \dfrac{1}{2\pi L_{y}}
  \int_{0}^{2\pi L_{x}} dx \int_{0}^{2\pi L_{y}} dy ~ \omega(x,y,t) e^{-ix/L_{x}}.
\end{equation}
We use its absolute value $|\wt{\omega}_{1,0}(t)|$ as the order parameter.
Three examples in the unstable side are demonstrated in Fig.~\ref{fig:FluidExample}
(note the different scales for the different curves).
We use three indicators to characterize the size of a pattern:
a time average $|\wt{\omega}_{1,0}|^{\rm av}$ for the codimension-two point, the first peak height $|\wt{\omega}_{1,0}|^{\rm fp}$ for the steady bifurcation side, and the maximum height $|\wt{\omega}_{1,0}|^{\rm max}$ for the oscillatory side; precise definitions are given below.

\subsubsection{Codimension-two discontinuous bifurcation}

\begin{figure}[h]
  \centering
    \includegraphics[width=8cm]{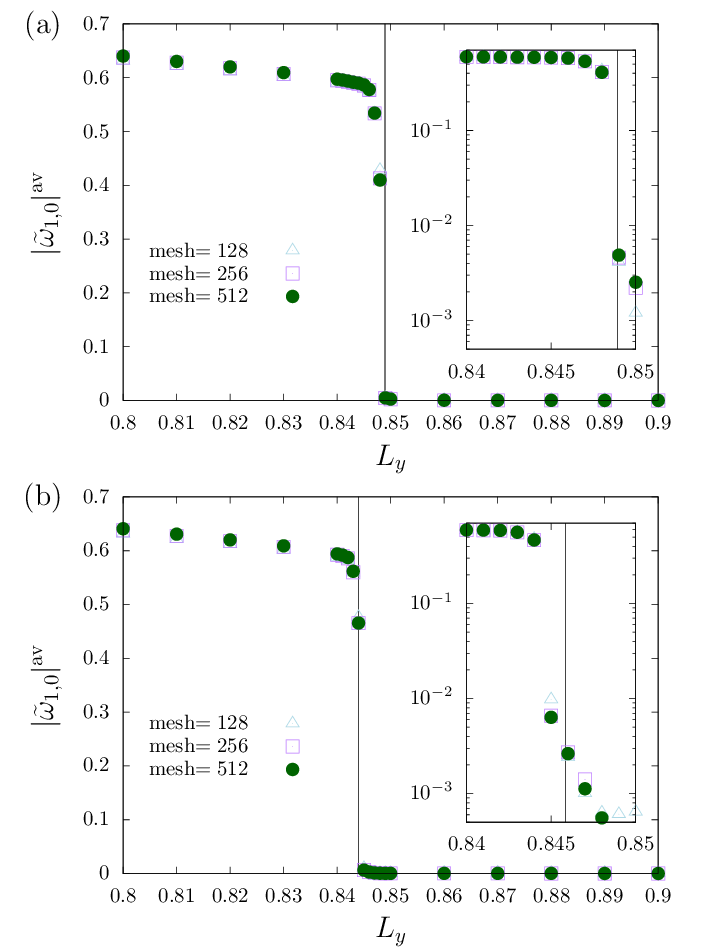}
  \caption{
    Discontinuous codimension-two bifurcation.
    The vertical axis represents the time average of $|\wt{\omega}_{1,0}|$.
    $(a_{2},a_{3}^{\rm col},L_{y}^{\rm c})$ is (a) $(0,0.0123,0.84892)$
    and (b) $(0.02, 0.0134, 0.84586)$. 
    These two points are on the line reported in the inset of Fig.~\ref{fig:FluidCriticalPoint}.
    The mesh size is 
    $M=128$ (blue triangles),
    $256$ (purple squares),
    and $512$ (green circles).
    The vertical straight lines indicate the critical points $L_{y}^{\rm c}$,
    which should be the discontinuous bifurcation points.
    Inset: Zoom around the predicted collision points.
    $\epsilon=10^{-3}$.
  }
  \label{fig:FluidDiscontinuity}
\end{figure}

To assess the existence of the codimension-two discontinuous bifurcation,
we monitor the time average
\begin{equation}
  |\wt{\omega}_{1,0}|^{\rm av}
  = \dfrac{1}{T_{2}-T_{1}} \int_{T_{1}}^{T_{2}} |\wt{\omega}_{1,0}(t)| dt,
\end{equation}
where $T_{1}=100$ and $T_{2}=300$.
We find in Fig.~\ref{fig:FluidDiscontinuity} a clear discontinuous bifurcation
at points where collisions at the origin occur.

In attractive Vlasov systems, around a codimension-two bifurcation point, the bifurcation is continuous, with trapping scaling, and closely followed by a jump in the order parameter, as recalled in Fig. \ref{fig:intro}. 
We now investigate the existence of this jump on both sides of the codimension two point.
For a given $a_2$, $a_{3}^{\rm col}$ corresponds to the eigenvalue collision at the origin, i.e. the codimension two point.
We call $a_{3}<a_{3}^{\rm col}$ ($\alpha<0$ on Fig.~\ref{fig:intro}) the steady bifurcation side (it corresponds to a critical eigenvalue crossing the imaginary axis at the origin, hence the critical mode is non oscillating), and $a_{3}>a_{3}^{\rm col}$ ($\alpha>0$ on Fig.~\ref{fig:intro}) the oscillatory bifurcation side (it corresponds to non zero complex conjugate eigenvalues crossing the imaginary axis, hence the critical mode is oscillating); see Fig.~\ref{fig:FluidEigenvalue} for an illustration.

\subsubsection{Steady bifurcation side}

\begin{figure}[ht]
  \centering
      \includegraphics[width=8cm]{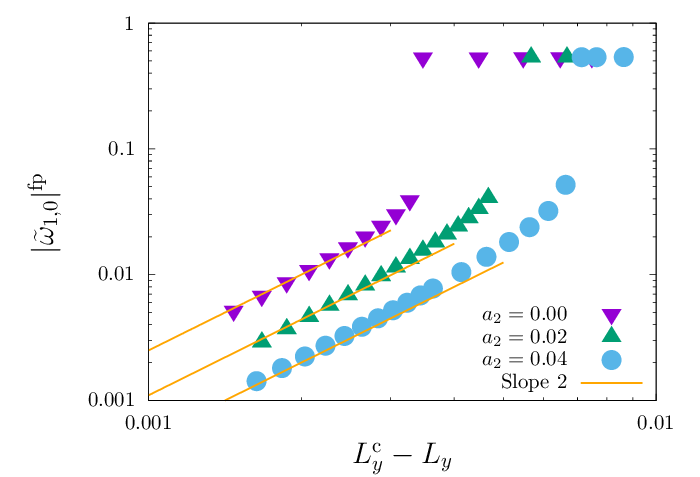}
    \caption{
      Trapping scaling and jump in the steady bifurcation side.
      The vertical axis $|\wt{\omega}_{1,0}|^{\rm fp}$ represents
      the first peak height of $|\wt{\omega}_{1,0}|$.
    $(a_{2},a_{3},L_{y}^{\rm c})$ is $(0, 0.011, 0.864471)$ (purple inverse triangles),
    $(0.02, 0.012, 0.862672)$ (green triangles),
    and $(0.04, 0.015, 0.854632)$ (blue circles),
    while the eigenvalue collision point is $(a_{2},a_{3})=(0,0.0123),~ (0.02,0.0134)$, and $(0.04, 0.0164)$.
    The mesh size is $M=256$.
    The orange straight lines have slope $2$.
  }
  \label{fig:FirstPeak}
\end{figure}

\begin{figure}[h]
  \centering
    \includegraphics[width=8cm]{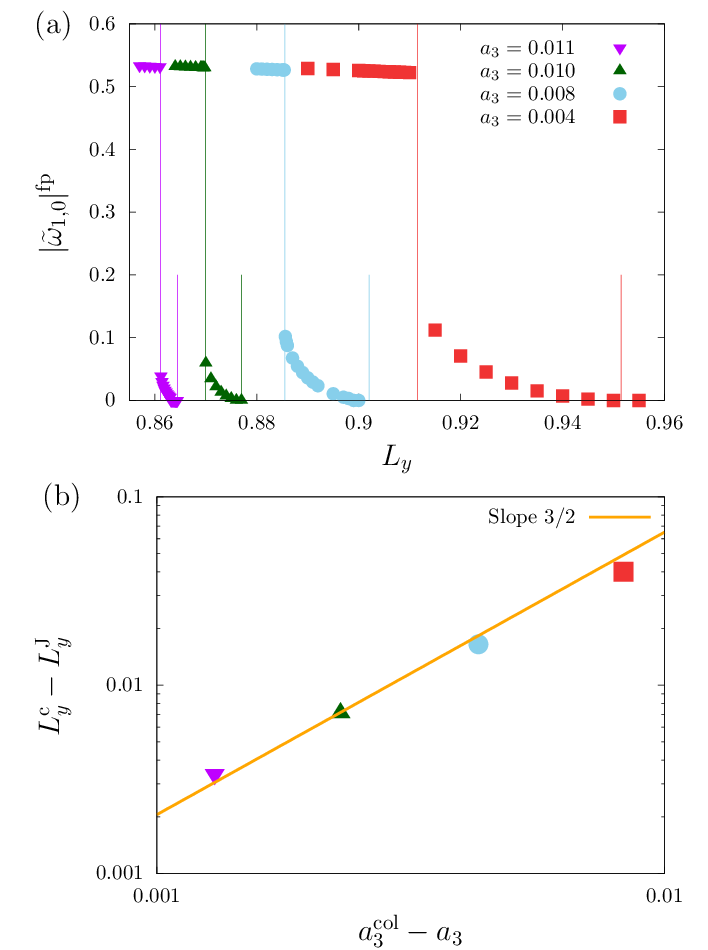}
  \caption{
    Jump scaling in the steady bifurcation side.
    (a) First peak height of $|\wt{\omega}_{1,0}(t)|$ for $a_{2}=0$.
    $a_{3}=0.011$ (purple inverse triangles),
    $0.010$ (green triangles),
    $0.008$ (blue circles),
    and $0.004$ (red squares)
    from left to right.
    The vertical short and long straight lines mark the critical
    and the jump points, respectively.
    (b) The distance between the critical point and the jump point, $L_{y}^{\rm c}-L_{y}^{\rm J}$
    as a function of the distance from the collision point, $-\Delta a_{3}=a_{3}^{\rm col}-a_{3}>0$.
    The orange straight line is a guide for the eyes and has slope $3/2$.
  }
  \label{fig:JumpScalingUnimodal}
\end{figure}

We fix three values of $a_{2}$: $0.00, 0.02$, and $0.04$.
Then, $a_{3}$ is chosen as $a_{3}=0.011$, $0.012$, and $0.015$ respectively,
which are smaller than $a_{3}^{\rm col}=0.0123, 0.0134$, and $0.0164$.
For a fixed pair of $(a_{2}, a_{3})$,
we compute the $L_{y}$ dependence of the first peak height of
$|\wt{\omega}_{1,0}(t)|$, denoted by $|\wt{\omega}_{1,0}|^{\rm fp}$.
The first peak height is reported in Fig.~\ref{fig:FirstPeak},
and we have two conclusions.
First, a jump does follow a continuous bifurcation for each $a_{2}$.
Second, the order parameter scaling right after the continuous bifurcation is $O((L_{y}^{\rm c}-L_{y})^{2})$,
which is equivalent to $O(\lambda^{2})$, i.e. trapping scaling.

We call $L_{y}^{\rm J}$ the length at which the order parameter jump occurs, and we compute the distance between the jump and the bifurcation: $L_{y}^{\rm c}-L_{y}^{\rm J}$. The scaling of this quantity is reported in Fig.~\ref{fig:JumpScalingUnimodal}. We numerically find an exponent $3/2$, which confirms the findings of \cite{Yamaguchi-Barre-23}.

\subsubsection{Oscillatory bifurcation side}
\begin{figure}[hbtp]
  \centering
    \includegraphics[width=8cm]{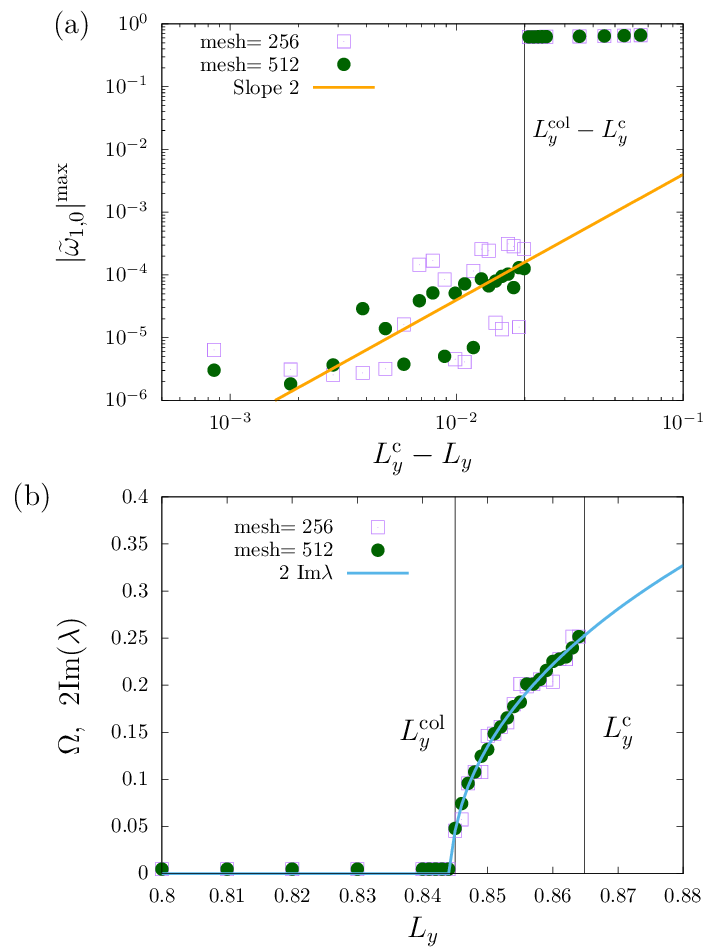}
  \caption{
    Trapping scaling and frequency in the oscillatory bifurcation side.
    (a) $|\wt{\omega}_{1,0}|^{\rm max}$, the maximum value of $|\wt{\omega}_{1,0}|$
    in the interval $t\in [0,3000]$.
    $(a_{2},a_{3})=(0, 0.0126)$.
    The critical point is $L_{y}^{\rm c}=0.865$
    and the eigenvalue collision point is $L_{y}^{\rm col}=0.845$.
    (b) Frequency of $|\wt{\omega}_{1,0}(t)|$ and $2~{\rm Im}(\lambda)$.
    The frequency is obtained as the peak position of the power spectrum
    for $\omega>0.003$.
    We removed points for $L_{y}>L_{y}^{\rm c}$ since no pattern is formed in these cases.
    The mesh size is $M=256$ (purple squares) and $512$ (green circles).
    }
  \label{fig:BimodalSidea20}
\end{figure}

The oscillatory bifurcation side requires higher precision.
We fix $a_{2}=0$ and choose $a_{3}=0.0126$,
which is greater than $a_{3}^{\rm col}=0.0123$.
We observe the maximum value of $|\wt{\omega}_{1,0}(t)|$ in the time interval $[0,3000]$,
denoted by $|\wt{\omega}_{1,0}|^{\rm max}$.
The maximum value is reported in Fig.~\ref{fig:BimodalSidea20}(a) and reveals three facts.
First, a jump follows a continuous bifurcation as in the steady bifurcation side. The jump point $L_{y}^{\rm J}$
is numerically indistinguishable from $L_{y}^{\rm col}$, the point where the two eigenvalues collide on the positive real axis, see Fig.~\ref{fig:FluidEigenvalue}.
Second, a continuous regime is sandwiched between the critical point $L_{y}^{\rm c}$
and $L_{y}^{\rm col}$ (or $L_{y}^{\rm J}$).
Third, the trapping scaling is unclear, but compatible with our simulations, which used up to $M=512$ mesh points.
A definite answer would require higher resolution, which we leave as a future work.

In the attractive Vlasov system in \cite{Yamaguchi-Barre-23}, the emergence of the jump is explained
by the merging of two clusters close to the eigenvalue collision. These clusters form at the frequency of the two unstable complex conjugate eigenvalues.
We can test this scenario by computing the frequency $\Omega$ of $|\wt{\omega}_{1,0}(t)|$.
There are four ``clusters'' in this case (two positive and two negative small vortices), and two each run in opposite directions with the same speed, hence the expected frequency is twice the imaginary part of the unstable eigenvalues.
Figure \ref{fig:BimodalSidea20}(b) convincingly confirms this scenario for the 2D Euler shear flow.

\begin{figure}[hbtp]
  \centering
    \includegraphics[width=8cm]{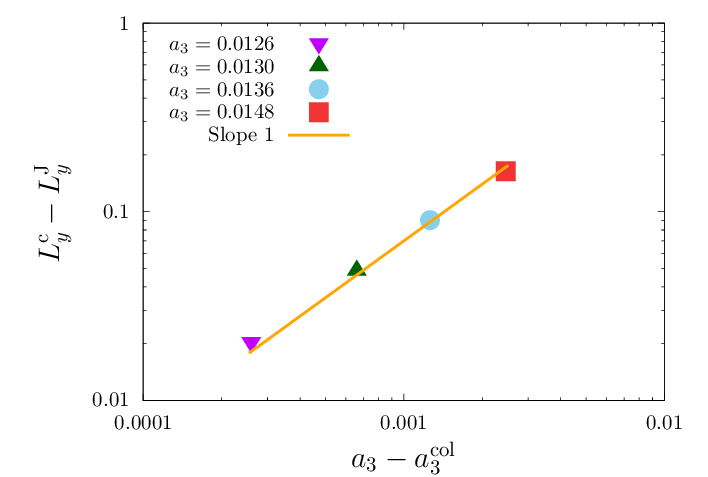}
  \caption{
    Jump scaling in the oscillatory bifurcation side.
     The distance between the critical point and the jump point, $L_{y}^{\rm c}-L_{y}^{\rm J}$
    as a function of the distance from the collision point, $\Delta a_{3}=a_{3}-a_{3}^{\rm col}>0$.
    The orange straight line has slope $1$.
    The mesh size is $M=128$.
  }
  \label{fig:JumpscalingBimodal}
\end{figure}

Finally, we investigate the behavior of the distance between the jump and the bifurcation: $L_{y}^{\rm c}-L_{y}^{\rm J}$.
Figure \ref{fig:JumpscalingBimodal} shows that this quantity scales as ${\rm Re}(\lambda)$, which again fully confirms the findings of \cite{Yamaguchi-Barre-23}.

\section{Repulsive case}
\label{sec:repulsive-case}

\subsection{Model}

We now consider a spatially one-dimensional periodic system.
The two-body interaction potential $\phi(q)$ is written in Fourier series
$\phi(q) = -\sum_{m\in \mathbb{N}} K_{m} \cos(mq)$.
For simplicity, we keep only the first two terms in the series:
\begin{equation}
  \phi(q) = -K_{1} \cos q - K_{2} \cos (2q),
\end{equation}
where $K_1,K_2<0$ to mimic a repulsive interaction. In the following $K_2=-1$ is fixed,
and $K_1$ is considered a bifurcation parameter.  
We call $F(q,p,t)$ the distribution function in position/momentum; Vlasov equation reads
\begin{eqnarray}
&&\partial_t F + p\partial_q F -\partial_q \Phi[F] \partial_p F = 0 \label{eq:Vlasova}\\
&&\Phi[F](q,t) = \iint F(q',p,t)\phi(q-q') dq' dp.\label{eq:Vlasovb}
\end{eqnarray}
Any spatially homogeneous distribution $F(p)$ is stationary for \eqref{eq:Vlasova}--\eqref{eq:Vlasovb}, and our goal is to investigate bifurcations from such stationary states.
We will detect bifurcations through the order parameter
\begin{equation}
  M_{1}(t) = \left| \int_{\mathbb{R}} dp \int_{-\pi}^{\pi} dq e^{iq} F(q,p,t) \right|.
\end{equation}
$M_{1}=0$ indicates a homogeneous phase in space, and $M_{1}>0$ a non homogeneous phase.
A unimodal symmetric momentum distribution is never unstable in a repulsive system,
hence we consider a family of bimodal distributions mimicking a two beams instability, with or without a small bump at zero velocity:
\begin{equation}
  F_{0}(p) = \dfrac{1-r}{2} \left[ G_{\sigma_{0}}(p-p_{0}) + G_{\sigma_{0}}(p+p_{0}) \right]
  + r G_{\sigma}(p),
  \label{eq:F-Plasma}
\end{equation}
where $0\leq r\leq 1$ and $p_{0},\sigma,\sigma_{0}>0$ are real parameters.
Note that $F_{0}(p)$ is normalized as $\int_{\mathbb{R}} F_{0}(p)dp=1$.
The function $G_{\sigma}(p)$ defined by
\begin{equation}
  G_{\sigma}(p) = \dfrac{1}{\sqrt{2\pi\sigma^{2}}} e^{-p^{2}/(2\sigma^{2})}
\end{equation}
is the normalized Gaussian distribution with zero-mean and standard deviation $\sigma$.
We will fix $\sigma_{0}=1$ and the family \eqref{eq:F-Plasma} has three parameters
$r, p_{0}$, and $\sigma$.
$F_{0}$ is flat around $p=0$, i.e. $F_{0}''(0)=0$, if $\sigma=\sigma^{\rm flat}(r,p_{0})$, where
\begin{equation}
  \sigma^{\rm flat}(r,p_{0})
  = \left[ \dfrac{r}{1-r} \dfrac{\sigma_{0}^{5}}{p_{0}^{2}-\sigma_{0}^{2}}
    e^{p_{0}^{2}/(2\sigma_{0}^{2})} \right]^{1/3}.
  \label{eq:flat-sigma}
\end{equation}
See Fig.\ref{fig:PlasmaEigenvalue}(a) for illustrations.

\subsection{Linear analysis}

We first compute the eigenvalues 
of the linearized Vlasov equation, close to the instability threshold.
We assume $K_{1}<K_{2}<0$, which ensures the instability is in the first Fourier mode.
The computation of the continued spectrum function of the first Fourier mode is classical, and it reads
\begin{equation}
  D_{1}(\lambda) = 1 + \dfrac{K_{1}}{2} \int_{\rm L} \dfrac{F_{0}'(p)}{p-i\lambda} dp,
\end{equation}
where the Landau integral contour ${\rm L}$ gives
\begin{equation}
  \int_{\rm L} \dfrac{F_{0}'(p)}{p-i\lambda} dp = \left\{
    \begin{array}{ll}
      \displaystyle{ \int_{\mathbb{R}} \dfrac{F_{0}'(p)}{p-i\lambda} dp } & ({\rm Re}\lambda>0), \\
      \displaystyle{ {\rm PV} \int_{\mathbb{R}} \dfrac{F_{0}'(p)}{p-i\lambda} dp + i\pi F_{0}'(i\lambda) } & ({\rm Re}\lambda=0), \\
      \displaystyle{ \int_{\mathbb{R}} \dfrac{F_{0}'(p)}{p-i\lambda} dp + i 2\pi F_{0}'(i\lambda) } & ({\rm Re}\lambda<0), \\
    \end{array}
  \right.
\end{equation}
and PV represents the Cauchy principal value integral. We need to solve $D_{1}(\lambda) =0$. Strictly speaking, roots with positive real part are eigenvalues, and roots with negative real part are Landau poles. We shall call all of them eigenvalues for simplicity.

We expect that the eigenvalue passes through the origin at the codimension-two bifurcation point which we are looking for.
Setting then $\lambda=0$, the critical point $K_{\rm c}$ is obtained from the equation
\begin{equation}
  1 + \dfrac{K_{\rm c}}{2}  {\rm PV} \int_{\mathbb{R}} \dfrac{F_{0}'(p)}{p} dp = 0.
\end{equation}
Since $K_{\rm c}<0$ (repulsive system), this imposes
\begin{equation}
  I(r,p_{0},\sigma) = {\rm PV} \int_{\mathbb{R}} \dfrac{F_{0}'(p)}{p} dp > 0.
  \label{eq:I}
\end{equation}
We take a set $(r,p_{0})$ so that $I(r,p_{0},\sigma^{\rm flat})>0$.
A representative example is chosen as $(r,p_{0})=(0.01,2)$,
which corresponds to $\sigma^{\rm flat}\simeq 0.2919$.
The critical point of the flat case is $K_{\rm c}\simeq -12.51$.
We remark that a small $r$ and a large $p_{0}$ is preferable to satisfy $I>0$;
otherwise $F_{0}(p)$ becomes close to a unimodal distribution.

\begin{figure}[htbp]
  \centering
      \includegraphics[width=9cm]{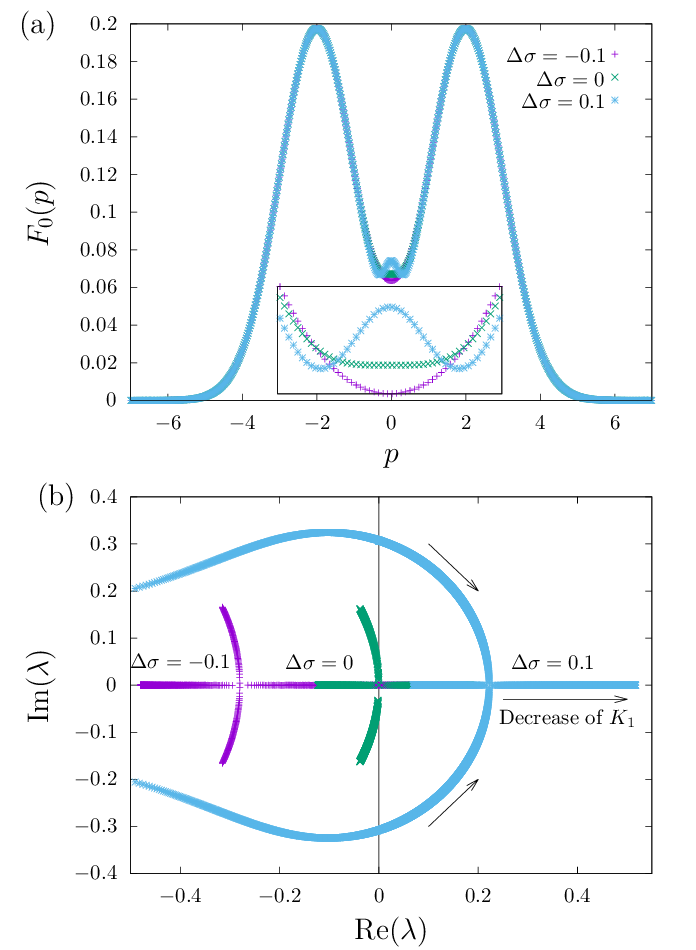}
    \caption{
      (a) Reference homogeneous distributions $F_{0}(p)$ [see Eq.\eqref{eq:F-Plasma}].
      $\Delta\sigma=\sigma^{\rm flat}-\sigma$.
      Inset: Zoom in the interval $p\in [-0.5,0.5]$.
      (b) Movement of eigenvalues, root of $D_{1}(\lambda)$,
      on the complex $\lambda$ plane.
      $(r,p_{0})=(0.01, 2)$, $\sigma^{\rm flat}\simeq 0.2919$, and $\sigma_{0}=1$.
  }
  \label{fig:PlasmaEigenvalue}
\end{figure}

We report on  Fig.~\ref{fig:PlasmaEigenvalue} the eigenvalues with largest real part,
as a function of $K_{1}$, varying $\sigma$ around $\sigma^{\rm flat}$.
For $\sigma>\sigma^{\rm flat}$, the bifurcation is non oscillatory: a real eigenvalue crosses the imaginary axis at the origin.
For  $\sigma<\sigma^{\rm flat}$, the bifurcation is oscillatory: at the critical point, two purely imaginary eigenvalues cross the imaginary axis.
For  $\sigma=\sigma^{\rm flat}$, i.e.  when the reference distribution $F_{0}(p)$
has a flat bottom at $p=0$, two eigenvalues collide precisely at the origin. $\Delta\sigma=\sigma^{\rm flat}-\sigma$ hence plays the role of the $\alpha$ parameter of Fig.~\ref{fig:intro}.

\subsection{Numerical tests}
The Vlasov equation is numerically integrated using the semi-Lagrangian algorithm
described in \cite{deBuyl-10} with the timestep $\Delta t=0.01$.
The phase space $(q,p)$ is truncated as $(q,p)\in [-\pi,\pi)\times [-7,7]$,
and the truncated phase space is divided in a $N\times 2N$ mesh.

\begin{figure}[htbp]
  \centering
  \includegraphics[width=9cm]{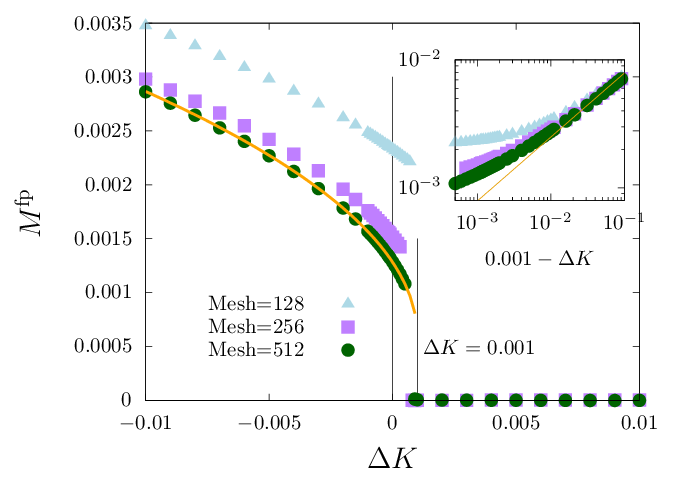}
  \caption{
    Codimension-two discontinuous bifurcation in a repulsive system.
    First peak height $M^{\rm fp}$ as a function of $\Delta K=K_{1}-K_{\rm c}$.
    $(r,p_{0},\sigma)=(0.01, 2, \sigma^{\rm flat})$ with $\sigma_{0}=1$.
    The flat distribution is realized following Eq.~\eqref{eq:flat-sigma}.
    The orange curve represents $M^{\rm fp}=0.0216\sqrt{0.001-\Delta K}+0.0006$,
    which suggests the existence of a very small jump.
    Inset: $M^{\rm fp}$ vs $0.001-\Delta K$ in logarithmic scale.}
  \label{fig:Bifurcation}
\end{figure}

\begin{figure}[htbp]
  \centering
  \includegraphics[width=9cm]{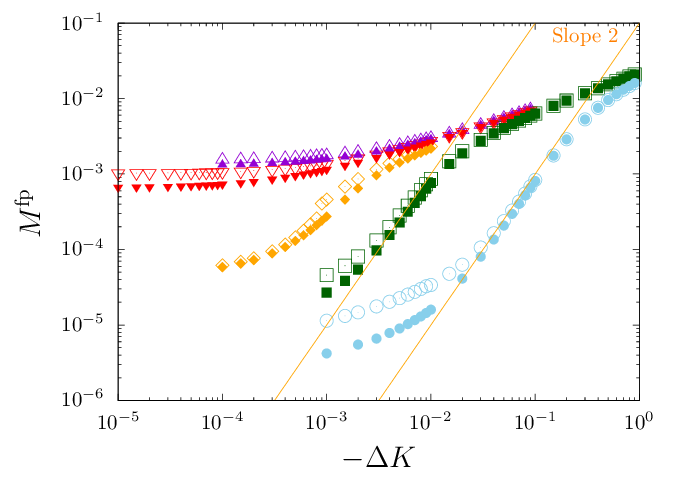}
  \caption{
    First peak height as a function of $-\Delta K=K_{\rm c}-K_{1}$.
    Steady bifurcation side.
    $(r,p_{0})=(0.01, 2)$ with $\sigma_{0}=1$.
    $\Delta\sigma=\sigma^{\rm flat}-\sigma=0$ (purple triangles),
    $-0.0001$ (red inverse triangles),
    $-0.0003$ (orange diamonds),
    $-0.001$ (green squares),
    and $-0.01$ (blue circles).
    Mesh size is $N=256$ (open symbols) and $N=512$ (filled symbols).
    Orange straight lines are guide of eyes with slope $2$.
  }
  \label{fig:Bifurcation-all}
\end{figure}

We observe the temporal evolution of $M_{1}(t)$
and compute its first peak height $M^{\rm fp}$ for the flat bottom distribution,
$(r,p_{0},\sigma)=(0.01,2,\sigma^{\rm flat})$ with $\sigma_{0}=1$.
For these parameters, different values of $K_1$ are scanned around the critical
value $K_c$. Hence $\Delta K := K_1-K_c =0$ corresponds to the expected codimension-two bifurcation point,
where two eigenvalues collide at the origin.
Figure~\ref{fig:Bifurcation} suggests that the bifurcation is discontinuous. 
However, the order parameter jump at the codimension-two bifurcation point is very small,
and a higher precision is required to reproduce the critical point precisely.
To reproduce the diagram in Fig. \ref{fig:intro}, for $\sigma \neq \sigma^{\rm flat}$ but close to it, we expect to see a region of continuous bifurcation with trapping scaling of the order parameter, followed by an order parameter jump when $|K_1|$ is increased further.  
Figure \ref{fig:Bifurcation-all} confirms the existence of a region with trapping scaling around $\sigma^{\rm flat}$, but due to lack of numerical precision, it 
is hard to reproduce the jump. A complete reproduction of the codimension-two bifurcation diagram is left as a future work.

\subsection{Size of the jump}

We discuss now why the jump is so small in this case. 
We base our discussion on the self-consistent equation
\cite{Yamaguchi-Barre-23,Leoncini09,deBuyl-11,OgawaYamaguchi14,OgawaYamaguchi15,TacuBenisti22},
a useful heuristic to understand the asymptotic state of the system in the vicinity of a bifurcation point.

At the critical point for the flat distribution, the self-consistent equation
gives (see appendix C)
\begin{equation}
  0 = L_{5/2} |K_{1}M|^{5/2} + L_{3} |K_{1}M|^{3} + \cdots,
  \label{eq:sce}
\end{equation}
where the $K_{1}$ factors appearing in $L_{5/2}$ and $L_{3}$ have been normalized to the unity,
by replacing $M$ with $K_{1}M$. The jump at the critical point is approximated by
\begin{equation}
  |M|^{\rm jump} = \left( - \dfrac{L_{5/2}}{\sqrt{|K_{\rm c}|} L_{3}} \right)^{2}
  \sim \dfrac{1}{|K_{\rm c}|} \left( \dfrac{F_{0}^{(4)}(0)}{\int_{\infty}^{\infty} [F_{0}^{(5)}(p)/p] dp} \right)^{2}
  \label{eq:jump-size}
\end{equation}
where $F_{0}^{(n)}$ is the $n$th derivative of $F_{0}$.
In our repulsive case the critical point is $|K_{\rm c}|\simeq 12.51$,
while the previous study in an attractive case gives $K_{\rm c}\simeq 0.97$.
Furthermore, the remaining factor coming from $F_{0}(p)$ also contributes to have a smaller estimation for $ |M|^{\rm jump}$.
In Appendix \ref{sec:JumpEstimation}, we show that  \eqref{eq:jump-size} is indeed roughly consistent with the very small numerical observation.

\section{Summary and conclusions}
\label{sec:summary}

We have tested the scenario of \cite{Yamaguchi-Barre-23} for the appearance of discontinuous bifurcations at a codimension-two point, 
both for 2D shear flows described by Euler equation and for repulsive particles systems described by Vlasov equation, mimicking a two-streams instability. The two systems are analyzed in the linear regime analytically and in the nonlinear regime numerically.
 
For 2D shear flows, we have been able to reproduce almost completely the scenario of  \cite{Yamaguchi-Barre-23}: the bifurcation is discontinuous when two eigenvalues collide at the origin; close to this codimension two point, the bifurcation is continuous, followed by a jump. The location of this jump follows a peculiar scaling, predicted in \cite{Yamaguchi-Barre-23} and again found here. The order parameter seems to follow a trapping scaling in the continuous region, as expected, although we have not been able to confirm this clearly on the oscillatory side of the codimension-two bifurcation. In addition, the shear flow case shows that a flat vorticity profile (the analog of a flat velocity distribution in the Vlasov case) is not necessary for this codimension-two bifurcation point. 

For the repulsive Vlasov equation, we have found the expected eigenvalue collision at the origin, and we have seen numerically a jump at this codimension two point. The jump is however very small, which precludes a more detailed numerical analysis, as in the shear flow case. The size of the jump could be increased by suitably choosing the reference homogeneous stationary state.

In both cases, and even when the discontinuity of the order parameter is very small, the codimension-two bifurcation point strongly limits in its neighborhood the validity of the trapping scaling, and hence of the Single Wave Model picture.
Many models with a collisionless dynamics similar to the ones studied in this article present possible discontinuous bifurcations, for instance \cite{Bonifacio84,Kaur18}; it would be interesting to see if they can be related to a codimension-two bifurcation point analogous to the one discussed in this paper.

\appendix

\section{Eigenvalue collision and flatness of reference state}
\label{sec:collision-flatness}

We show the equivalence between a collision of two eigenvalues at the origin
and the flatness of the reference even momentum distribution $F_{0}(p)$ in a one-dimensional spatially periodic Vlasov system.
The analytically continued spectrum function for the Fourier mode $k~(k\neq 0)$ is
\begin{equation}
  D_{k}(\lambda) = 1 + \dfrac{K_{k}}{2} \int_{\rm L} \dfrac{F_{0}'(p)}{p-i\lambda/k} dp,
\end{equation}
where the integral contour ${\rm L}$ is the real axis for ${\rm Re}(\lambda)>0$
but is continuously modified to avoid the singularity at $p=i\lambda/k$ for ${\rm Re}(\lambda)\leq 0$.
The constant $K_{k}$ represents the coupling strength in the Fourier mode $k$,
and we assume that $K_{k}\neq 0$ as a necessary condition to have instability.
The normalization condition is $\int_{\mathbb{R}} F_{0}(p) dp=1$.

The collision of two eigenvalues implies that $D_{k}(\lambda)$ has a double
root at $\lambda=0$, hence $D_{k}(0)=D_{k}'(0)=0$.
The first condition $D_{k}(0)=0$ determines the critical value for $K_{k}$,
and we consider the second condition $D_{k}'(0)=0$.
Performing the integration by parts, the derivative is
\begin{equation}
  D_{k}'(\lambda) = \dfrac{iK_{k}}{2k} \int_{\rm L} \dfrac{F_{0}''(p)}{p-i\lambda/k} dp.
\end{equation}
For $\lambda=0$, the modification of the integral contour gives half the residue contribution at $p=0$
and the remaining principal value part vanishes by symmetry. Therefore, we have
\begin{equation}
  D_{k}'(0) = \dfrac{iK_{k}}{2k} i\pi \,{\rm sgn}(k) F_{0}''(0),
\end{equation}
and the second condition $D_{k}'(0)=0$ is equivalent with $F_{0}''(0)=0$.

\section{Semi-Lagrangian method}
\label{sec:semi-Lagrangian}

The Euler equation is expressed by using the total derivative as
\begin{equation}
  \dfrac{D\omega}{Dt}
  = \dfracp{\omega}{t} + \dfracp{\psi}{y} \dfracp{\omega}{x}
  - \dfracp{\psi}{x} \dfracp{\omega}{y} = 0;
\end{equation}
the value of $\omega$ is then conserved along a solution $(x(t),y(t))$
to the Hamiltonian equations of motion
\begin{equation}
  \label{eq:EOM}
  \dot{x} = u(x,y) = \dfracp{\psi}{y}, \qquad \dot{y} = v(x,y) = -\dfracp{\psi}{x}.
\end{equation}
A semi-Lagrangian method uses the evolution relation
from the time $t-\Delta t$ to $t$:
\begin{equation}
  \omega(x(t), y(t), t)
  = \omega(x(t-\Delta t), y(t-\Delta t), t-\Delta t).
\end{equation}

Numerically, we divide the $(x,y)$ plane into a lattice,
and keep the values of $\omega$ only at the lattice points.
Therefore, $(x(t),y(t))$ should be a lattice point.
Hence the basic building blocks of a semi-Lagrangian method are 
a backward evolution according to Eq.~\eqref{eq:EOM} from a lattice point $(x_{m},y_{n})=(x(t),y(t))$
to $(x(t-\Delta t),y(t-\Delta t))$, which is not a lattice point in general,
and an interpolation of $\omega$ to get $\omega(x(t-\Delta t),y(t-\Delta t),t-\Delta t)$.
See \cite{deBuyl-10} for a one-dimensional separable Hamiltonian system,
which allows us to use an explicit symplectic integrator
and a one-dimensional interpolation.
The Hamiltonian $\psi$ in the two-dimensional Euler is not separable in general,
and we use other algorithms to realize the two building blocks.

\subsection{Temporal evolution}

Since the Hamiltonian $\psi$ is not separable in general,
we use the Taylor series expansion with respect to the time $t$.
The point $(x(t-\Delta t),y(t-\Delta t))$ is obtained as
\begin{equation}
  \begin{split}
    & x(t-\Delta t) = x(t) - \Delta t~u(x,y) \\
    & + \dfrac{\Delta t^{2}}{2} \left[
      \dfracp{u}{x}(x,y) u(x,y) + \dfracp{u}{y}(x,y) v(x,y)
      \right] + O(\Delta t^{3}), \\
    & y(t-\Delta t) = y(t) - \Delta t~v(x,y) \\
    & + \dfrac{\Delta t^{2}}{2} \left[
      \dfracp{v}{x}(x,y) u(x,y) + \dfracp{v}{y}(x,y) v(x,y)
      \right] + O(\Delta t^{3}),
  \end{split}
\end{equation}
where we used the equations of motion \eqref{eq:EOM} repeatedly,
and $(x,y)$ is evaluated at the time $t$ and at a lattice point in the right-hand side.
We remark that a second order Runge-Kutta algorithm requires an interpolation
to obtain the values of $u$ and $v$ at a non-lattice point.

The velocity field and its derivatives are obtained
using the fast Fourier transform (FFT).
We write the Fourier series expansion of the stream function $\psi(x,y)$, for instance, as
\begin{equation}
  \psi(x,y) = \sum_{k,l} \wt{\psi}_{k,l} e^{ikx/L_{x}} e^{ily/L_{y}}.
\end{equation}
The relation $\omega=-\nabla^{2}\psi$ implies the relation for the $(k,l)$-Fourier components
of $\wt{\psi}_{k,l}$ and $\wt{\omega}_{k,l}$:
\begin{equation}
  \wt{\psi}_{k,l} = \left\{
    \begin{array}{ll}
      0 & (k,l) = (0,0) \\
      \dfrac{\wt{\omega}_{k,l}}{k^{2}/L_{x}^{2}+l^{2}/L_{y}^{2}} & \text{otherwise.} \\
    \end{array}
  \right.
\end{equation}
The physically meaningless constant $\wt{\psi}_{0,0}$ is assumed to be zero.
Noting that $u, v, \partial_{x}u, \partial_{y}u, \partial_{x}v$,
and $\partial_{y}v(=-\partial_{x}u)$ are expressed as derivatives of $\psi$,
their Fourier components are obtained through the multiplication of $\wt{\psi}_{k,l}$
by $ik/L_{x}$ (resp. $il/L_{y}$) for the derivative
$\partial_{x}$ (resp. $\partial_{y}$).
Finally, the values of $u$ at the lattice points, for instance,
are obtained by performing the inverse FFT of $\{\wt{u}_{k,l}\}$.

\subsection{Bicubic interpolation}

Let the lattice spacings be $\Delta x$ and $\Delta y$
for the $x$ and $y$ axes respectively, and $(x_{0},y_{0})$ be a lattice point.
A lattice point $(x_{m},y_{n})$ is defined by $(x_{0}+m\Delta x, y_{0}+n\Delta y)$, where $m,n\in\mathbb{Z}$.
We change the variables as
\begin{equation}
  \bar{x} = \dfrac{x-x_{0}}{\Delta x},
  \quad
  \bar{y} = \dfrac{y-y_{0}}{\Delta y},
\end{equation}
and map a lattice point $(x_{m},y_{n})$ to an integer lattice point $(\bar{x}_{m},\bar{y}_{n})=(m,n)$.
By shifting the indices, we may assume without loss of generality
that the inversely evolved point $(x(t-\Delta t), y(t-\Delta t))$
is on a rectangle $[x_{0},x_{1}]\times [y_{0},y_{1}]$, which is mapped to $[0,1]\times [0,1]$.

\begin{figure}[hbp]
  \centering
  \includegraphics[width=6cm]{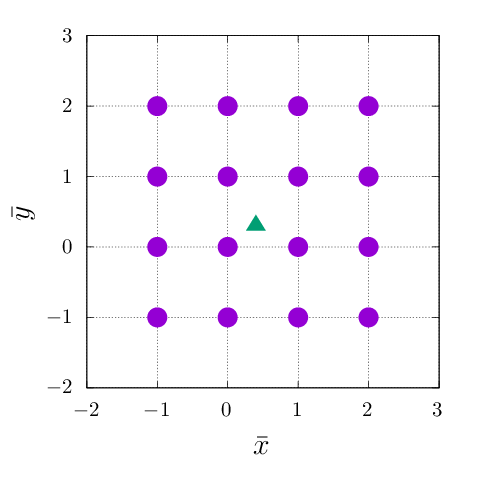}
  \caption{
    Schematic picture for bicubic interpolation on the scaled plane $(\bar{x},\bar{y})$.
    An interpolation at a target point (green triangle) is performed
    by using 16 lattice points (purple circles).
  }
  \label{fig:FluidSemiLagrangian}
\end{figure}

For simplicity of notation, we denote the scaled variables $(\bar{x},\bar{y})$ by $(x,y)$
and perform an interpolation at a point $(x,y)\in [0,1]\times [0,1]$
for a function $f(x,y)$ whose values are known only on a two-dimensional integer lattice.
The key idea is to approximate $f(x,y)$ by a bicubic function
\begin{equation}
  \begin{split}
    p(x,y)
    & = \sum_{i,j=0}^{3} a_{ij} x^{i} y^{j}
    =
      \begin{pmatrix}
        1 & x & x^{2} & x^{3}
      \end{pmatrix}
      A
      \begin{pmatrix}
        1 \\ y \\ y^{2} \\ y^{3} \\
      \end{pmatrix},
  \end{split}
\end{equation}
where the matrix $A$ is
\begin{equation}
  A = 
  \begin{pmatrix}
    a_{00} & a_{01} & a_{02} & a_{03} \\
    a_{10} & a_{11} & a_{12} & a_{13} \\
    a_{20} & a_{21} & a_{22} & a_{23} \\
    a_{30} & a_{31} & a_{32} & a_{33} \\
  \end{pmatrix}.
\end{equation}
To determine the $16$ coefficients $a_{ij}$,
we choose $16$ lattice points around the rectangle $[0,1]\times [0,1]$ as $(m,n)~(m,n\in\{-1,0,1,2\})$
(see Fig.~\ref{fig:FluidSemiLagrangian}).
Requiring the equality $f(m,n)=p(m,n)$ for all the $16$ lattice points, we have the relation
\begin{equation}
  F = C^{\rm T} A C,
\end{equation}
where
\begin{equation}
  F = 
  \begin{pmatrix}
    f(-1,-1) & f(-1,0) & f(-1,1) & f(-1,2) \\
    f(0,-1) & f(0,0) & f(0,1) & f(0,2) \\ 
    f(1,-1) & f(1,0) & f(1,1) & f(1,2) \\ 
    f(2,-1) & f(2,0) & f(2,1) & f(2,2) \\ 
  \end{pmatrix},
\end{equation}
$C^{\rm T}$ is the transpose of $C$, and 
\begin{equation}
  C =
  \begin{pmatrix}
    1 & 1 & 1 & 1 \\
    -1 & 0 & 1 & 2 \\
    1 & 0 & 1 & 4 \\
    -1 & 0 & 1 & 8 \\
  \end{pmatrix}.
\end{equation}
The coefficient matrix $A$ is obtained as
\begin{equation}
  A = C^{-{\rm T}} F C^{-1},
\end{equation}
where $C^{-{\rm T}}$ is the transpose of $C^{-1}$ and
\begin{equation}
  C^{-1} = \dfrac{1}{6}
  \begin{pmatrix}
    0 & -2 & 3 & -1 \\
    6 & -3 & -6 & 3 \\
    0 & 6 & 3 & -3 \\
    0 & -1 & 0 & 1 \\
  \end{pmatrix}.
\end{equation}

The above algorithm used $16$ lattice points of $9$ small rectangles to interpolate $f(x,y)$
at $(x,y)\in [0,1]\times [0,1]$,
but an interpolation can be performed by using only $4$ lattice points of $1$ small rectangle,
if we use the values of $f$, $\partial_x f$, $\partial_yf$, and $\partial_{xy}f$ at the $4$ lattice points.
On a periodic two-dimensional lattice, the Fourier coefficient $(\wt{\partial_{xy}f})_{k,l}$
is obtained by multiplying $\wt{f}_{k,l}$ by a second order term in $k$ and $l$,
and $\partial_{xy}f$ is obtained from the inverse FFT of $\wt{\partial_{xy}f}$.
However, this multiplication induces large errors for large $k$ or $l$,
and it is better to use the values of $f$ on the extended region
$(x,y)\in[-1,2]\times[-1,2]$.

\section{Self-consistent equation}
\label{sec:SCE}

The self-consistent equation
\cite{Yamaguchi-Barre-23,Leoncini09,deBuyl-11,OgawaYamaguchi14,OgawaYamaguchi15,TacuBenisti22}
predicts the final stationary state $F^{\rm fin}(q,p)$ from the initial state $F_{0}(p)$.
Let us denote the final one-body Hamiltonian by
\begin{equation}
  H^{\rm fin}(q,p) = \dfrac{p^{2}}{2} + V(q),
\end{equation}
where $V(q)$ is the one-body potential. For instance, in the Hamiltonian mean-field (HMF) model,
the potential is
\begin{equation}
  V(q) = - K_{1} M \cos q,
\end{equation}
where $M$ is defined by
\begin{equation}
  M = \int_{\mathbb{R}} dp \int_{0}^{2\pi} dq~ \cos q F^{\rm fin}(q,p).
\end{equation}
we performed a suitable rotation to satisfy $\int_{\mathbb{R}}dp\int_{0}^{2\pi} dq~\sin q \,F^{\rm fin}(q,p)=0$
without loss of generality.
In the repulsive HMF system, we have $K_{1}<0$ and $M<0$,
and the final Hamiltonian is read as
\begin{equation}
  H^{\rm fin}(q,p) = \dfrac{p^{2}}{2} - |K_{1}| |M| \cos q.
\end{equation}
This is formally the same expression as the attractive HMF model,
and we replace $K_{1}$ and $M$ with $|K_{1}|$ and $|M|$ in the following.

The final state $F^{\rm fin}(q,p)$ is assumed to depend on $(q,p)$ only through the final Hamiltonian,
which makes it stationary for the Vlasov equation.
We note that the final Hamiltonian $H^{\rm fin}(q,p)$ is integrable,
so that we can introduce angle-action variables $(\theta,J)$.
The self-consistent theory predicts the final state $F^{\rm fin}(q,p)$ as
\begin{equation}
  F^{\rm fin}(q,p) = \ave{F_{0}(p)}_{J},
  \label{eq:SC-distribution}
\end{equation}
where $\ave{\cdots}_{J}$ represents the average over the angle variable,
in other words along a contour with constant action variable $J$,
associated with the final Hamiltonian $H^{\rm fin}(q,p)$.
Both sides of \eqref{eq:SC-distribution} depend on the final state,
and it must be solved self-consistently.

Recalling that $M<0$ and that we replace $M$ with $|M|$,
an expansion of \eqref{eq:SC-distribution} gives the self-consistent equation
for $|M|$ as
\begin{equation}
  - D_{1}(0) |M| =  L_{3/2} |M|^{3/2} + L_{5/2} |M|^{5/2} + L_{3} |M|^{3} + \cdots.
  \label{eq:SC}
\end{equation}
Coefficients are
\begin{equation}
  \begin{split}
     L_{3/2}|M|^{3/2} &= \dfrac{F^{(2)}(0)}{2} \iint p^{2} \left(
      \ave{\cos q}_{J} - \dfrac{c_{1}(q)}{p^{2}} \right) dqdp \\
     L_{5/2} |M|^{5/2} &= \dfrac{F^{(4)}(0)}{4!} \iint p^{4} \left(
      \ave{\cos q}_{J} - \dfrac{c_{1}(q)}{p^{2}} - \right.\\
      &\left. \dfrac{c_{2}(q)}{p^{4}} \right) dqdp \\
     L_{3} |M|^{3} &= \dfrac{1}{5!} \int c_{3}(q) dq \int \dfrac{F^{(5)}(p)}{p} dp
  \end{split}
\end{equation}
and $c_{n}(q)$'s are defined by
\begin{equation}
  \ave{ \cos q}_{J} = \sum_{n=1}^{\infty} \dfrac{c_{n}(q)}{p^{2n}}
  \quad
  ( |p| \to\infty).
\end{equation}
A remarkable difference with the attractive case is the sign of the left-hand side
in \eqref{eq:SC}.
This opposite sign requires that the coefficients on the right hand side also have opposite signs in order
to keep the same scenario as in the attractive case \cite{Yamaguchi-Barre-23}:
$F^{(4)}(0)$, for instance, must be negative in an attractive system,
but it must be positive in a repulsive system.

The right-hand side of \eqref{eq:SC-distribution} is defined by $H^{\rm fin}(q,p)$,
which depends on $|K_{1}M|$, and the coefficients depend on $K_{1}$.
We can eliminate the $K_{1}$ dependence by replacing $|M|$ with $|K_{1}M|$
in the right-hand side of \eqref{eq:SC}. Ensuring the conditions for the codimension two points are met leads to Eq.~\eqref{eq:sce}.

\section{Estimation of the jump size}
\label{sec:JumpEstimation}

For the repulsive system investigated in Sec.~\ref{sec:repulsive-case},
we use the parameters $(r,p_{0},\sigma_{0})=(0.01,2,1)$ and the corresponding
flat distribution with $\sigma^{\rm flat}\simeq 0.2919$.
They give the values 
\begin{equation}
  F_{0}^{(4)}(0) \simeq 5.37748,
  \qquad
  \int_{\mathbb{R}} \dfrac{F_{0}^{(5)}(p)}{p} dp \simeq -128.773.
\end{equation}
Since the critical point is $K_{\rm c}\simeq -12.51$, an estimation of the jump size is
\begin{equation}
  |M|^{\rm jump} \simeq \dfrac{1}{12.51} \left( \dfrac{5.37748}{-128.773} \right)^{2}
  \simeq 1.394\times 10^{-4}.
  \label{eq:MjumpRepulsive}
\end{equation}
This estimation is smaller than the observed jump, which is approximately $10^{-3}$
(see Fig.~\ref{fig:Bifurcation}).
Recall that we neglected contribution from $c_{n}$ factors.
We will compare this estimation with the attractive case investigated in
the previous work \cite{Yamaguchi-Barre-23},
under the assumption that the contributions from the $c_n$ factors are of the same order of magnitude in both cases.

In the attractive system the reference homogeneous distribution was
\begin{equation}
  F_{\rm att}(p) = C \exp\left[ -\beta_{2}p^{2}/2 - (\beta_{4}p^{2}/2)^{2} \right],
  \quad
  \beta_{4}=3.
\end{equation}
The flat distribution is realized with $\beta_{2}=0$
and the normalization factor $C$ is in this case $C=\Gamma(1/4)/\sqrt{2\beta_{4}}$,
where $\Gamma(z)$ is the gamma function defined by
\begin{equation}
  \Gamma(z) = \int_{0}^{\infty} t^{z-1} e^{-t} dt.
\end{equation}
For this flat distribution we obtain
\begin{equation}
  F_{\rm att}^{(4)}(0) = -6C\beta_{4}^{2},
  \quad
  \int_{-\infty}^{\infty} \dfrac{F_{\rm att}^{(5)}(p)}{p} dp
  = 24 C \beta_{4}^{2}\sqrt{2\beta_{4}} \Gamma(3/4).
\end{equation}
With the critical point $K_{\rm c}\simeq 0.97$, an estimation of the jump size is
\begin{equation}
  |M|^{\rm jump}_{\rm att}
  \simeq \dfrac{1}{0.97} \left( \dfrac{1}{4\sqrt{2\beta_{4}}\Gamma(3/4)} \right)^{2}
  \simeq 0.00715.
  \label{eq:MjumpAttractive}
\end{equation}
Again, this estimation is smaller than the observed jump, which is approximately $0.1$.

Now we have estimations of jump sizes for the repulsive case \eqref{eq:MjumpRepulsive}
and for the attractive case \eqref{eq:MjumpAttractive}.
The ratio between the two estimations is 
\begin{equation}
  \dfrac{|M|^{\rm jump}}{|M|^{\rm jump}_{\rm att}}
  \simeq 0.0195
\end{equation}
and is of the same order as the observed ratio $0.01$.

\end{document}